\documentclass{article}
\newdimen\footheight	
\usepackage{sprocl,graphicx}

\newcommand{\eg}{e.g.\ }
\newcommand{\ie}{i.e.\ }
\newcommand\FC{\textit{FormCalc}}
\newcommand\cuba{\textsc{Cuba}}

\begin{document}

\hfill {\small MPP--2004--97}

\title{New Developments for Automatic Loop Calculations}

\author{Thomas Hahn}

\address{Max-Planck-Institut f\"ur Physik \\
        F\"ohringer Ring 6 \\
	D--80805 Munich, Germany}

\maketitle

\abstracts{Two methods to reduce the CPU time needed for the numerical
evaluation of cross-sections and similar quantities are discussed.}


\section{Introduction}

The numerical evaluation of cross-sections and similar quantities can be
expensive in terms of CPU time, in particular if radiative correction
are included, and/or if the model has parameters that need to be scanned
over.  This contribution discusses new and improved algorithms for
phase-space integration and parallelization of parameter scans to speed
up the calculation.  Both methods have successfully been implemented in
version 4 of the package \FC\ \cite{FormCalc}.


\section{Phase-space integration}

The recently completed \cuba\ library \cite{Cuba} provides four
subroutines for multidimensional numerical integration.  All four have a
very similar invocation and can thus be interchanged easily, \eg for
comparison.  The flexibility of a general-purpose method is particularly
useful in the setting of automatically generated code.  The following 
algorithms are contained in the \cuba\ library:

\textit{Vegas} is the classic Monte Carlo algorithm which uses
importance sampling for variance reduction.  It iteratively builds up a
piecewise constant weight function, represented on a rectangular grid. 
Each iteration consists of a sampling step followed by a refinement of
the grid.  The present implementation uses Sobol quasi-random numbers
for sampling.

\textit{Suave} is a crossover between Vegas and Miser and combines
Vegas-style importance sampling with globally adaptive subdivision:
Until the requested accuracy is reached, the region with the largest
error is bisected along the axis in which the fluctuations of the
integrand are reduced most.  In each half the number of new samples is 
prorated for the fluctuation.

\textit{Divonne} is a further development of the \textsc{CernLib}
routine D151.  It is intrinsically a Monte Carlo algorithm but has
cubature rules built in for comparison, too.  The variance-reduction
method is stratified sampling.  In a first step, a tessellation of the
integration region is constructed in which all subregions have an
approximately equal value of the spread, defined as
\begin{equation}
s(r) = \frac 12\mathop{\mathrm{Vol}}(r)
   \Bigl(\max_{\vec x\in r} f(\vec x\,) -
         \min_{\vec x\in r} f(\vec x\,)\Bigr).
\end{equation}
Minimum and maximum here are sought using methods from numerical
optimization.  The subregions are then sampled independently with a
number of points extrapolated to reach the required accuracy.  For each
region, the latterly obtained value is compared to the initial rough
estimate and if the two are not compatible within their errors, the
region is subdivided or sampled once more.  Additions to 
\textsc{CernLib}'s D151 are the final comparison phase and the
possibility to point out known extrema, to speed up convergence.

\textit{Cuhre} is a new implementation of \textsc{dcuhre}.  It is a
deterministic algorithm which employs cubature rules of a polynomial
degree.  Variance reduction is by globally adaptive subdivision: Until
the requested accuracy is reached, bisect the region with the largest
error along the axis with the largest fourth difference.

\begin{figure}
\centerline{\includegraphics[width=.9\linewidth]{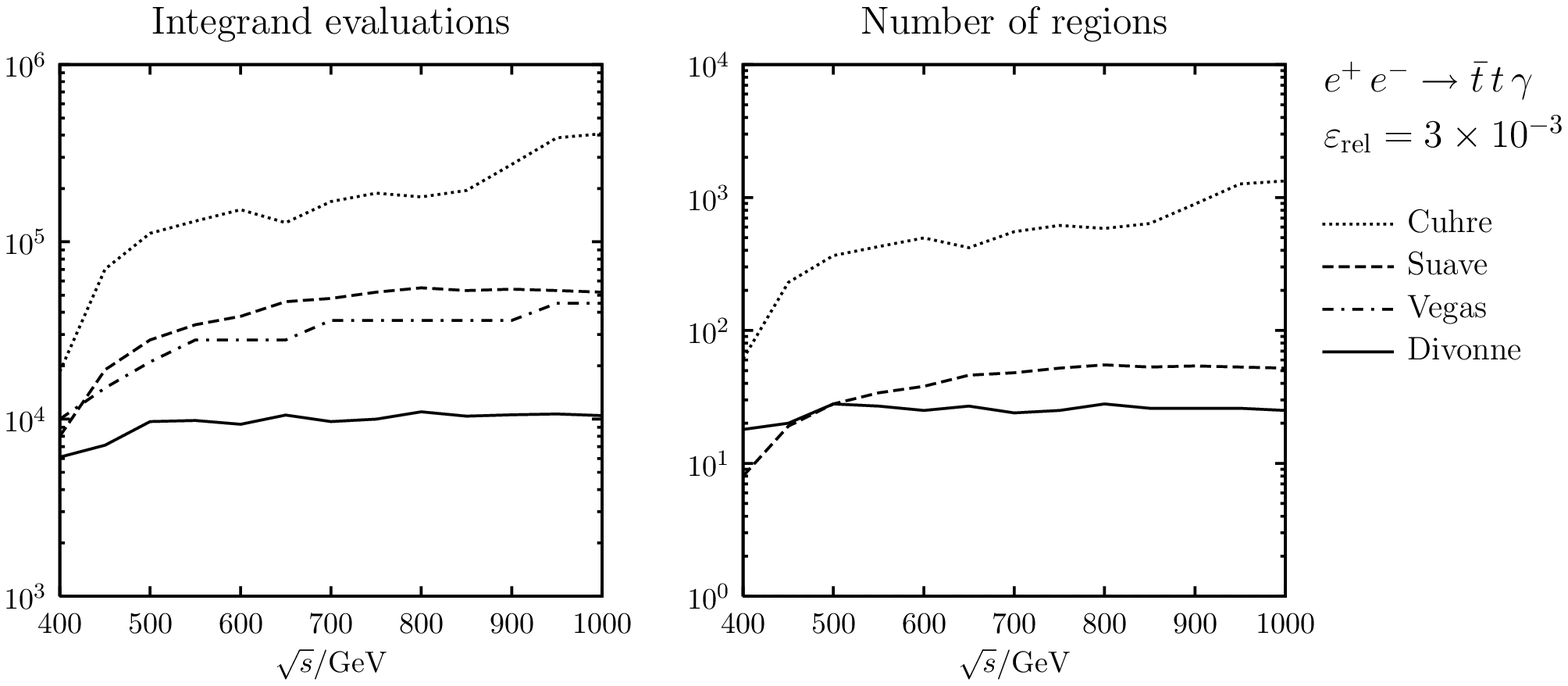}}
\caption{\label{fig:eettA}Characteristics of the \cuba\ routines
for the phase-space integration of $e^+e^-\to\bar t t\gamma$ at a 
requested relative accuracy of $3\times 10^{-3}$.}
\end{figure}

Fig.\ \ref{fig:eettA} compares the performance of the four algorithms
for a real phase-space integration of the process $e^+e^-\to\bar t
t\gamma$.  Above all it is very important to have several independent
integration methods to cross-check the results.

\section{Parallelization}

Calculations in models like the MSSM, where not all input parameters are
yet known, often require extensive scans to cover an interesting part of
the parameter space.  Such a scan can be a real CPU hog, but on the
other hand, the calculation can be performed completely independently
for each parameter set and is thus an ideal candidate for
parallelization.  The real question is thus not how to parallelize the
calculation, but how to automate the parallelization.

The method is quite general, but consider \FC\ for a specific instance. 
The user may specify parameter loops by defining preprocessor variables,
\eg
\begin{verbatim}
   #define LOOP1 do 1 TB = 2, 30
\end{verbatim}
These definitions are substituted at compile time into a main loop 
(see below).  The obstacle to automatic parallelization is that the 
loops are user-defined and in general nested.  A serial number is 
introduced to unroll the loops:

\smallskip

\begin{center}
\begin{small}
\begin{tabular}{|l|l|l|}
\cline{1-1}\cline{3-3}
\textit{serial version} && \textit{parallel version} \\
\cline{1-1}\cline{3-3}
			&& \verb|   serial = 0| \\
\verb|   LOOP1|		&& \verb|   LOOP1| \\
\verb|   LOOP2|		&& \verb|   LOOP2| \\
\verb|     |$\vdots$	&& \verb|     |$\vdots$ \\
			&& \verb|   serial = serial + 1| \\
			&& \verb|   if( serial |\textit{not in allowed range}\verb| ) goto 1| \\
\verb|   |\textit{calculate cross-section}\quad
			&& \verb|   |\textit{calculate cross-section} \\
\verb|1  continue|	&& \verb|1  continue| \\
\cline{1-1}\cline{3-3}
\end{tabular}
\end{small}
\end{center}

\smallskip

The serial number range can be specified on the command line so that it
is quite straightforward to distribute patches of serial numbers on
different machines.  Most easily this is done in an interleaved manner,
since one then does not need to know to which upper limit the serial
number runs, \ie if there are $N$ machines available, send serial
numbers 1, $N + 1$, $2N + 1$, etc.\ on machine 1, send serial numbers 2,
$N + 2$, $2N + 2$, etc.\ on machine 2, \dots

This procedure is completely automated in \FC: The user once creates a
\texttt{.submitrc} file in his home directory and lists there all
machines that may be used, one on each line.  In the case of
multi-processor machines he puts the number of processors after the host
name.  The executable compiled from \FC\ code, typically called
\texttt{run}, is then simply prefixed with \texttt{submit}.  For
instance, instead of ``\texttt{run uuuu 500,1000}'' the user invokes
``\texttt{submit run uuuu 500,1000}.''  The \texttt{submit} script uses
\texttt{ruptime} to determine the load of the machines and \texttt{ssh}
to log in.  Handling of the serial number is invisible to the user.



\newcommand{\volyearpage}[3]{\textbf{#1} (#2) #3}
\newcommand{\cpc}{\textsl{Comp.\ Phys.\ Commun.} \volyearpage}
\newcommand{\npps}{\textsl{Nucl.\ Phys.\ Proc.\ Suppl.} \volyearpage}

\begin{flushleft}

\end{flushleft}

\end{document}